**On the features of great Forbush effect during May 2024 extreme geomagnetic storm**

*M. A. Abunina[1], N. S. Shlyk[2], A. V. Belov[3], S. M. Belov[4], A. A. Abunin[5]*

*Pushkov Institute of Terrestrial Magnetism, Ionosphere and Radio Wave Propagation of Russian Academy of Sciences (IZMIRAN), 108840 Russia, Moscow, Troitsk, Kaluzhskoe Hw., 4*

**Abstract**

The work investigates the features of galactic cosmic ray density and anisotropy behavior and their relation to solar sources, interplanetary and geomagnetic disturbances from May 8 to May 13, 2024. During this time, powerful solar flares and fast CMEs were recorded, leading to registration of an extreme geomagnetic storm along with one of the most significant Forbush effects for the entire observation period. All the calculations of cosmic ray characteristics are made using the data of global neutron monitor network and unique methods maintained at IZMIRAN: the Global Survey Method and the Ring of Stations Method.

It is determined that the magnitude of Forbush effect under study was 15.7% (for particles with 10 GV rigidity) and as an extreme geomagnetic storm was recorded there was a significant magnetospheric effect observed in the data of neutron monitors (~4%).

1. **Introduction**

The most extreme solar activity events always prove to be of the highest interest to the scientific community. One such event occurred during May 2024, an outstanding burst of solar activity, which caused an extreme geomagnetic storm with record parameters. Several researchers have already published their works regarding this event. For instance, Hajra et al. [2024] discussed interplanetary causes and impacts of May Superstorm on the Geosphere. The authors thoroughly describe the behavior of interplanetary medium parameters, geomagnetic activity (GA) indices, ionospheric current flows, and cosmic rays (CR) variations according to some ground neutron monitor (NM) stations.

Liu et al. [2024] compared the "geo-effectiveness" of May storm based on data from satellites at the Earth's orbit and the Stereo-A spacecraft. In particular, authors demonstrate a significant disparity in the observed parameters, because the Earth and Stereo-A were situated in different regions of the interplanetary disturbance: "two contrasting cases of complex ejecta are found in terms of the geo-effectiveness at the Earth, which is largely due to different magnetic field configurations within the same active region".

In addition to providing a detailed description of the solar sources and interplanetary parameters of the events, Hayakawa et al. [2024] also gathered auroral records, reconstructed the

---

equatorward boundary of the visual auroral oval, and compared naked-eye and camera auroral visibility. A significant Forbush decrease and GLE 74 (Ground Level Enhancement, https://gle.oulu.fi, Usoskin et al., 2020), registered in specific ground detectors' data were described.

May 2024 events are also discussed in the work of Mavromichalaki et al. (2024), where the authors evaluate the efficiency of Ap-index forecasting methods and GLE alert system performance (likewise using single NM data).

At IZMIRAN, an advanced method (global survey method – GSM) for studying CR density variations was maintained. It's based on data from all available ground detectors (www.nmdb.eu) and allows one to obtain 1-hour data on CR flux variations observed by the Earth as a single multi-directional detector [Belov et al., 2018]. In particular, it is applicable for CR anisotropy study (the first harmonic magnitude and phase), which is almost impossible when using NM data directly.

In addition, IZMIRAN has a regularly updated database of Forbush effects and interplanetary disturbances - FEID (www.izmiran.ru/tools/feid). It contains 1-hour values of parameters of interplanetary medium, cosmic rays (with 10 GV rigidity) and geomagnetic activity (excluding Kp-indices with 3-hour resolution). Using the FEID it is possible to compare different characteristics of new events with the ones of previous solar cycles (since 1957) and make conclusions about what place they occupy among the most powerful and interesting events.

The aim of this work is to investigate the features of CR density and anisotropy behavior and their relation to solar sources and geomagnetic storms of the extreme events during May 2024. CR parameters are calculated with the GSM for 10 GV particles. The Ring of Stations method (RSM) is used as an alternative for CR anisotropy assessment. Both methods are based on the global neutron monitor network data.

## 2. Data availability

- FEID database - www.izmiran.ru/tools/feid
- NMDB - www.nmdb.eu
- IGLED - https://gle.oulu.fi
- Wind spacecraft data - https://space.umd.edu/pm/
- SDO data - https://sdo.gsfc.nasa.gov/data
- Coronal mass ejections - https://cdaw.gsfc.nasa.gov/CME_list
- Coronal holes - https://solen.info/solar/coronal_holes.html
- Solar flares and active regions - https://solarmonitor.org/, https://www.swpc.noaa.gov/products/solar-and-geophysical-event-reports
- Geomagnetic activity data - http://wdc.kugi.kyoto-u.ac.jp/dstdir/index.html, ftp://ftp.gfz-potsdam.de/pub/home/obs/kp-ap/wdc (Matzka et al., 2021)
- Proton flux, GOES - https://www.swpc.noaa.gov/products/goes-proton-flux
- Shockvawes - https://space.umd.edu/pm/
- SSC list - https://isgi.unistra.fr/events_sc.php

## 3. Results and discussion

*3.1 Solar activity, interplanetary medium, and geomagnetic activity*

On April 30, 2024, an active region (AR) 13664 appeared at the south-east limb of the solar disc. Later, between May 2 and May 15, it produced 89 significant X-ray flares, 48 ones belonging to M-class and 10 - to X-class (https://solarmonitor.org/, https://www.swpc.noaa.gov/products/solar-and-geophysical-event-reports)

In particular, on May 8, 11 flares were observed in this AR, with 3 of them being especially notable: X1.0 at 4:37 UT, M8.7 at 11:26, and X1.0 at 21:08. These flares were associated with coronal mass ejections (CMEs) with initial speeds of 530, 677 and 952 km/s respectively (here and further linear speeds are used, https://cdaw.gsfc.nasa.gov/CME_list/). Additionally, at ~18 UT, a big filament erupted from north-east of the central region (partial halo CME at 19:12 UT, V0 = 401 km/s). And lastly, the coronal hole CH1221 (positive polarity, https://solen.info/solar/coronal_holes.html) crossed the central solar meridian in the northern hemisphere. Figure 1 depicts the solar disc as observed on May 8, 2024 by the SDO spacecraft (https://sdo.gsfc.nasa.gov/data), including labels of aforementioned coronal holes and ARs (https://solarmonitor.org/).

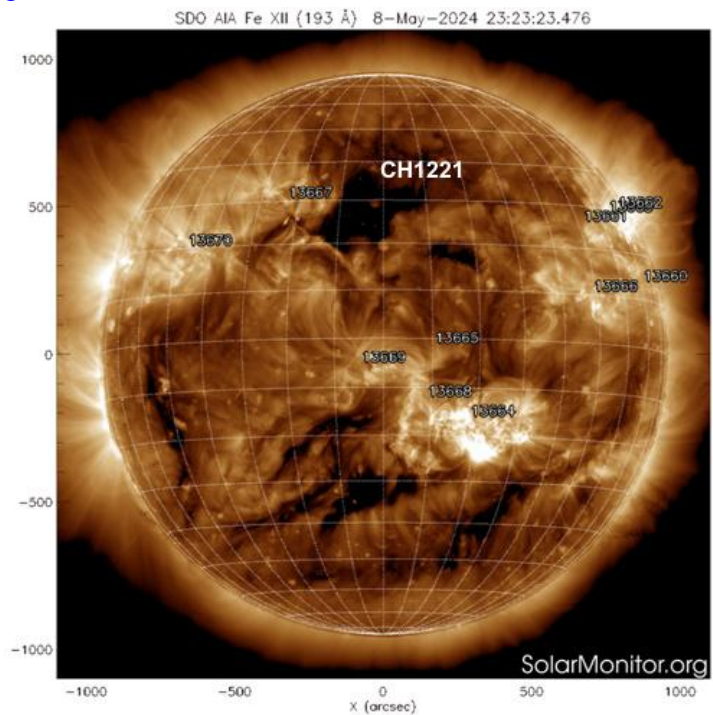

Figure 1: Solar disc observed on May 8 by SDO mission.

Activity in AR13664 did not stop further and 11 more M- and X-flares were registered on May 9. Including X2.2 (at 08:45 UT) and X1.1 (at 17:23 UT) flares, accompanied by halo CMEs with initial velocities of 1280 km/s and 1024 km/s respectively.

On May 10 and 11, the active region produced two more significant flares (X3.9 and X5.8), with fast CMEs, which lead to an increase in observed proton fluxes according to GOES data (https://www.swpc.noaa.gov/products/goes-proton-flux). The halo CME associated with the X3.9 flare was detected on May 10 at 07:12 UT with 953 km/s initial linear velocity. And another halo CME was registered with an initial speed of 1614 km/s on May 11 at 01:36 UT. Figure 2 shows the GOES data on particle fluxes with different energies in the upper panel, CME height-time plots in the middle one, and soft X-ray flux on the bottom (adopted from

https://cdaw.gsfc.nasa.gov/CME_list/daily_plots/sephtx). Incidentally, in the evening of May 10 an eruption of a big filament was observed in the south of the solar disc (partial halo CME at 21:17 UT, 686 km/s), which also made its contribution, complicating the interplanetary environment even further. All CMEs that were significant (to the authors' opinion) for near-Earth space, are labeled with numbers on the middle panel of Figure 2. For convenience, solar source data are summed up in Table 1.

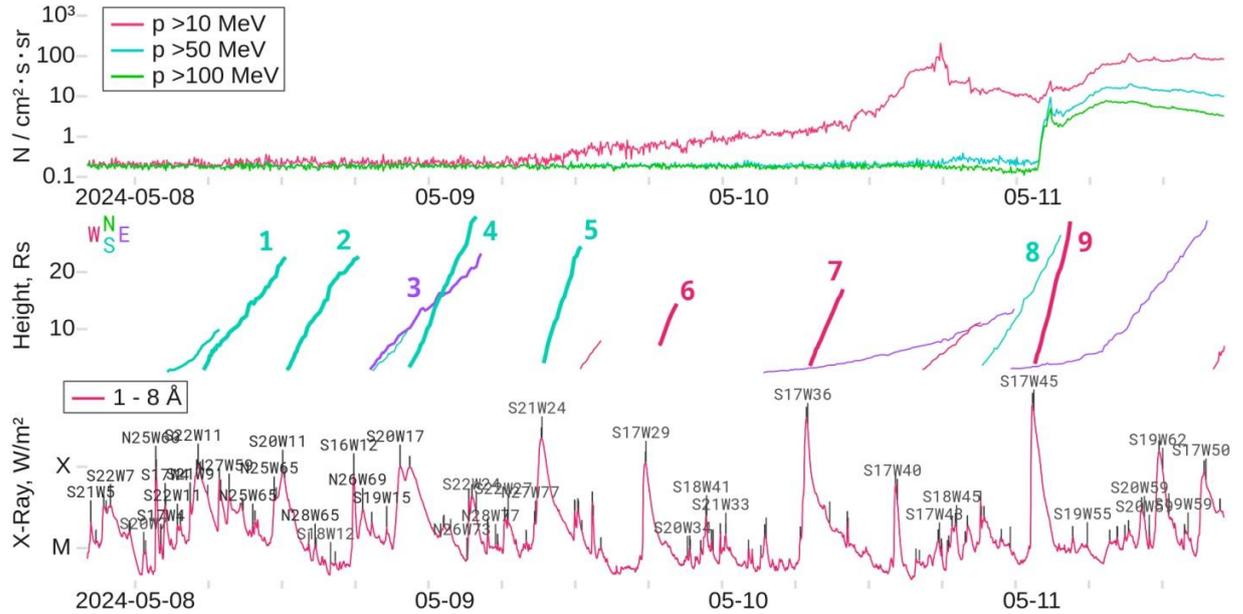

Figure 2. Proton flux, CME height-time and X-ray flux during May 8-11, 2024 (adopted from https://cdaw.gsfc.nasa.gov/CME_list/daily_plots/sephtx/).

Table 1. Solar sources of interplanetary disturbances under consideration.

| № | X-ray flare / DSF time, UT | X-ray flare Class | X-ray flare coordinates / DSF position | CME time, UT | CME linear speed, km/s | CME type |
|---|---|---|---|---|---|---|
| 1. | 2024.05.08 04:37 | X1.0 | S18W17 | 2024.05.08 05:36 | 530 | Halo |
| 2. | 2024.05.08 11:26 | M8.7 | S20W11 | 2024.05.08 12:24 | 677 | Halo |
| 3. | 2024.05.08 ~18:00 | DSF* | North-east of the center | 2024.05.08 19:12 | 401 | Partial Halo |
| 4. | 2024.05.08 21:08 | X1.0 | S20W17 | 2024.05.08 22:24 | 952 | Halo |
| 5. | 2024.05.09 08:45 | X2.2 | S20W26 | 2024.05.09 09:24 | 1280 | Halo |
| 6. | 2024.05.09 17:23 | X1.1 | S14W28 | 2024.05.09 18:52 | 1024 | Halo |
| 7. | 2024.05.10 06:27 | X3.9 | S17W34 | 2024.05.10 07:12 | 953 | Halo |
| 8. | 2024.05.10 ~20:00 | DSF* | South of the center | 2024.05.10 21:17 | 686 | Partial Halo |
| 9. | 2024.05.11 01:10 | X5.8 | S15W45 | 2024.05.11 01:36 | 1614 | Halo |

*DSF - disappeared solar filament

Aforementioned events on the Sun led to a complicated state of the interplanetary medium, observed near Earth between May 10 and May 13 (see Figure 3, upper panel). It aligns well with modeling results of several research groups (e.g. https://www.swpc.noaa.gov/products/wsa-enlil-solar-wind-prediction; https://iswa.gsfc.nasa.gov/downloads/20240511_060000_2.0_anim.tim-den.gif).

On May 10 at 16:38 UT an interplanetary shock was registered near Earth (according to Wind spacecraft data, https://space.umd.edu/pm/). It was a result of several interplanetary counterparts of CMEs on May 8-9. The fast CME of May 9 (CME №5) was, likely, the main driving force. It overran the structures from slower CMEs which came before it (CMEs №1–4 on May 8), interacting with all of them in the process.

In several hours (at ~21:00 UT), evidently, the Earth was impacted by another ICME, associated with a halo CME (№6 on May 9). As a result of all mentioned disturbances' influence, the parameters of interplanetary medium reached peak values. Solar wind (SW) speed, in particular, increased up to 800 km/s, and interplanetary magnetic field (IMF) strength almost hit 70 nT. Such magnitude of the IMF induction was never registered before in the period of instrumental observations, the last record was set on November 6, 2001 at 62 nT. SW plasma density also reached a significant value of 48.1 particles/cm$^3$ (Fig. 3, medium panel).

In the afternoon of May 11 (~17:00 UT), an ICME from another halo CME reached the Earth (CME №7 on May 10). Consequently, SW speed increased further, reaching 960 km/s. And in the afternoon of the next day (May 12), the Earth's magnetosphere was, apparently, impacted by ICMEs from the remainder of the described CMEs (CMEs №8, 9 on May 10, 11). Note that the high SW velocity on May 13 may be explained by an additional influence of a high-speed stream (HSS) from CH1221. The extreme values of SW, IMF, and GA parameters during the described interplanetary disturbances are also presented in Table 2.

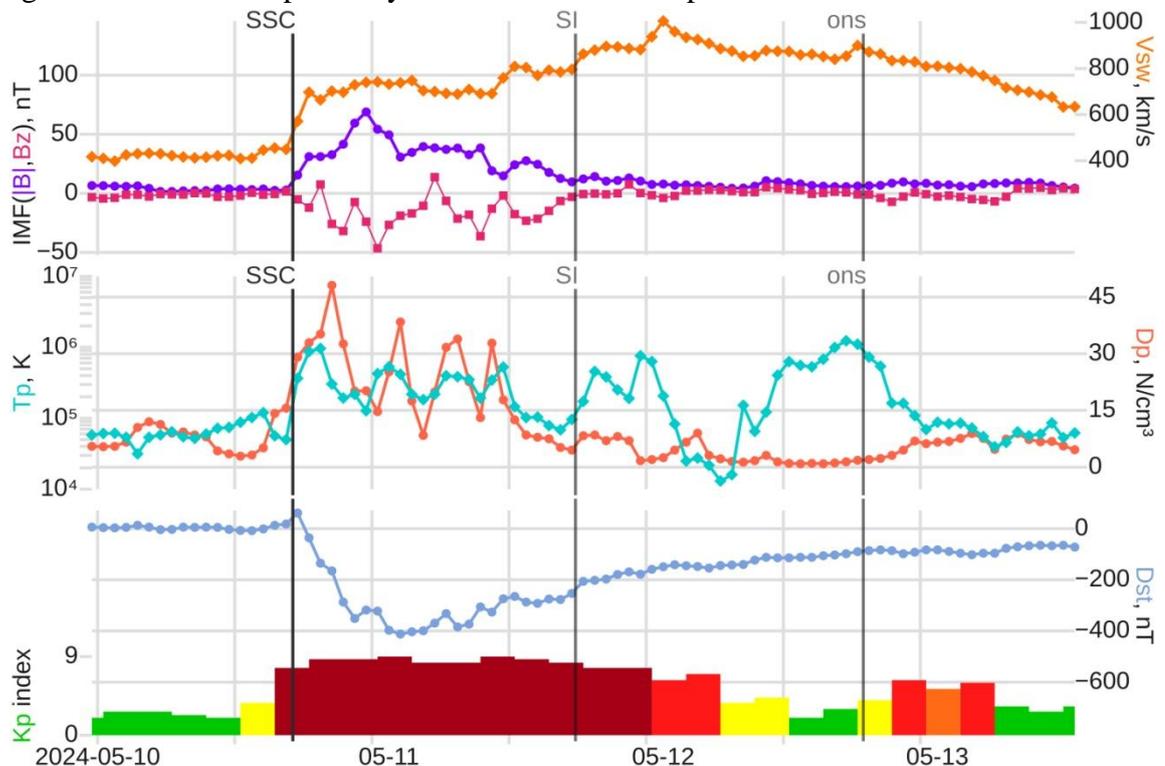

Figure 3. Interplanetary medium parameters and geomagnetic activity data on May 10-13, 2024 (upper panel: SW speed, IMF strength, IMF Bz-component; medium panel: SW density and temperature; lower panel: Dst- and Kp-indices of GA).

A sudden storm commencement (SSC) was registered at 17:05 UT on May 10. A 3-hour Kp-index reached a value of 8 and didn't go below 6 until 06:00 UT May 12. Such a long lasting geomagnetic storm is explained by impacts of several strong ICMEs described above, and is particularly related to the almost-record value of Bz = –47 nT (the absolute minimum is –54 nT, registered on November 6, 2001). During May 11 several intervals of extreme geomagnetic storm (Kpmax = 9) were observed; the last time comparable values were registered during the famous Halloween events of October 29-30, 2003 (e.g., Panasyuk et al., 2004; Belov, Eroshenko, and Yanke, 2005; Ermolaev et al., 2005; Gopalswamy et al., 2005; Tsurutani et al., 2006).

Dst-index of GA reached a minimum value of –412 nT (on May 11), which is also one of the record values since the beginning of space era, the other ones being observed on: March 14, 1989 (Dstmin = –589 nT); July 15, 1959 (Dstmin = –429 nT); September 13, 1957 (Dstmin = –427 nT); February 11, 1958 (Dstmin = –426 nT); and November 20, 2003 (Dstmin = –422 nT). The other prominent parameter is daily-averaged Ap-index, it reached a value of 271 on May 11, and a higher value was only registered once - on November 13, 1960 (Apd = 280). During the mentioned Halloween events, the daily-averaged Ap-index was equal to 204 and 191 on October 29 and 30 respectively.

Another geomagnetic storm (with Kpmax = 6) was observed from 21:00 UT May 12 to 06:00 UT May 13, being caused by a joint influence of ICMEs from CMEs №8-9 and an HSS from CH1221.

Table 2. Extreme values of the main SW, IMF, and GA parameters during storms on May 10 to May 13, 2024.

| № | Event onset time, UT | Vmax, km/s | Dmax, particles/cm$^3$ | Bmax, nT | Bzmin, nT | Dstmin, nT | Kpmax |
|---|---|---|---|---|---|---|---|
| 1 | 10.05.2024 17:05 | 814 | 48.1 | 69.8 | –40 | –412 | 9 |
| 2 | 11.05.2024 17:49 | 963 | 9 | 14.1 | –3.4 | –253 | 8+ |
| 3 | 12.05.2024 19:00 | 859 | 8.9 | 9.6 | –6.3 | –102 | 6+ |

### 3.2 Cosmic rays behavior

Publicly available database - the FEID (https://tools.izmiran.ru/feid) – contains data on CR parameters along with other interplanetary medium parameters during interplanetary disturbances and Forbush-effects (FEs) associated with them; we believe every interplanetary disturbance does modulate the CR flux to some extent. Hence, an FE is registered every time there is an abrupt change in SW and/or IMF parameters, or an SSC is detected. Oftentimes, events arrive one after each other, meaning that the beginning of one marks the "end" of the previous one. So here and further we consider an FE to be a change in anisotropy and density of cosmic ray flux under the influence of any interplanetary disturbance (Belov et al., 2001). During the described period of time (May 10 to 13, 2024) three interplanetary disturbances of a mixed nature were registered near Earth and there are three corresponding FEs in our database (see also Table 3): FE №1 starting at 17:05 UT, May 10 (https://isgi.unistra.fr/events_sc.php) associated with several CMEs of May 8-9; FE №2 starting at 17:49 UT, May 11 (https://space.umd.edu/pm/figs/figs.html), associated with CME №7; and FE №3 at 19:00 UT, May 12 associated with CMEs №8–9 and an HSS from CH1221. Start times of these FEs match either the SSC time or the time of interplanetary disturbance arrival. The main subject for the cosmic rays behavior analysis will be FE №1 as the other two were observed while CR flux did not yet fully recover from the first one. Main CR characteristics of these FEs calculated with the GSM for 10 GV particles are provided in Table 3. These parameters are thoroughly discussed below.

Table 3. Main FE parameters for 10 GV particles (calculated with the GSM) during May 10-13, 2024

| № | FE onset time, UT | FE magnitude, % | Dmin, % per hour | Axy max, % | Axy mean, % | Az range, % | Main solar source |
|---|---|---|---|---|---|---|---|
| 1 | 10.05.2024 17:05 | 15.7 | –4.4 | 1.91 | 0.9 | 3.1 | CMEs №1-6 |
| 2 | 11.05.2024 17:49 | 1.1 | –0.6 | 1.72 | 1.1 | 1.6 | CME №7 |
| 3 | 12.05.2024 19:00 | 2.8 | –0.4 | 2.45 | 1.1 | 3.3 | CMEs №8–9, HSS from CH1221 |

Additionally, the Ring of Stations method (RSM) was developed at IZMIRAN and implemented as a web application (https://tools.izmiran.ru/ros, Abunina et al., 2020). This method uses data of several specifically selected NM stations. Stations are selected in a way that allows using them together to obtain information on CR angular distribution in the Earth's ecliptic plane without advanced data processing. This method produces longitude-time plots, which may often show precursory signs of FEs 2-12 hours prior to the event onset. May 2024 events show precursor signs too. Particularly, a pre-increase at around 7 hours before the event onset (a group of blue circles visible in Figure 4 in a longitude range of 150-300° before the first SSC). This phenomenon has been also described in Mavromichalaki et al. (2024), so we will not go into much detail here. But we shall notice one peculiarity in the event development: the Forbush effect (in the form of NMs count rates decrease) starts only 2-3 hours after the SSC, which is apparent in all stations data (see Figure 4, magenta circles). This is not rare, yet most of the significant FEs do start immediately after an SSC (see, e.g., Abunina et al., 2020). Apparently, in the first hours, the magnetic connection between near-Earth space and the region outside the FE is still maintained, and, in addition, it may be explained by particles acceleration in front of the shockwave initially compensating for the usual CR depletion behind the front of the interplanetary disturbance.

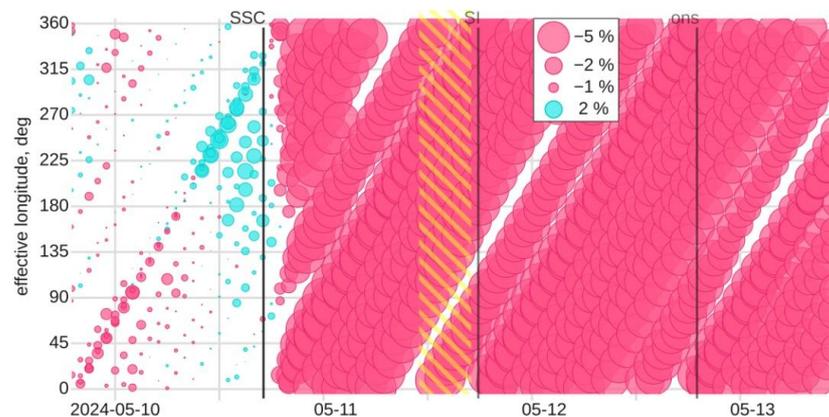

Figure 4. Longitude-time distribution of CR variation (based on the RSM) during May 10 – 13, 2024.

According to global NM network data processed with the GSM, the FE on May 10-11, 2024 turned out to be the biggest one for the last 20 years: its magnitude (for 10 GV particles) was found to be 15.7% (see Figure 5). The previous very large FE with magnitude of ~26% was registered on October 29, 2003, it still holds the record for all the period of observations.

Let's discuss the development of FE on May 10-11 in more detail. For the first two hours after the onset only a minimal decrease (~1%) was observed in the CR density. In the next two hours we observed a further decrease of ~3.8%, and then CR density dropped at an even greater pace of ~4.4% per hour for 2 hours, reaching its minimum at ~01:00 UT, May 11, after which the CR flux began to gradually recover. When compared by extreme parameters with other significant events, the event on May 10-11 does stand out. Greater values of minimum hourly decrement of CR density (Dmin) were observed only twice: on October 29, 2003 (–5.6% per hour) and on October 28, 1991 (–4.6% per hour). But when compared by decrement over two hours, a greater value was only observed in the second aforementioned event (–10.2%).

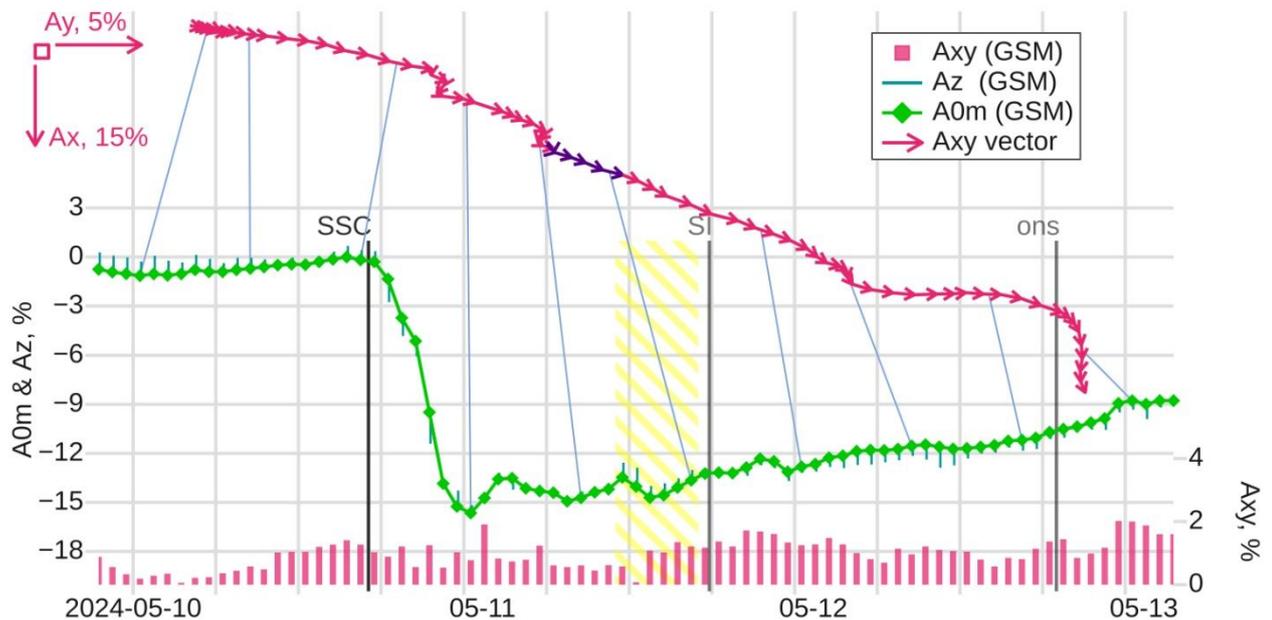

Figure 5. CR anisotropy and density (computed with the GSM for 10 GV particles on data of the global NM network) during May 10 –13, 2024.

As seen in Figure 5, the behavior of CR anisotropy is relatively calm yet interesting. The north-south anisotropy component variation range ($Az_{range}$) during FE №1 was equal to ~3.1% (see also Table 3), which is significantly lower than the average $Az_{range}$ = 4.26±0.3% calculated for all FEs with magnitude >10% recorded in the FEID database. The value of the CR anisotropy equatorial component ($Axy$) was quite low as well, the maximum value $Axy_{max}$ = 1.9% in the event is the minimum of all the FEs with magnitude >10% (the mean $Axy_{max}$ = 4.4±0.4%). The average value of equatorial anisotropy variation in this event $Axy$ = 0.9% is also minimum (the mean $Axy$ =1.7±0.1% for the FEs of >10% magnitude). Such abnormally small values of CR anisotropy may be related to the fact that at the time of discussed event in addition to radial CR gradient, latitudinal and/or azimuthal gradients were also present, indicating that the Earth was not located in the center of FE area, where the absolute minimum of CR density would be observed, but as bit to the west of it. It means that the absolute minimum of this FE should have been significantly lower than the 15.7% observed near the Earth. Suggested configuration could also explain the aforementioned "delayed onset" of the FE (apparent in Figure 4).

CR flux density did not recover to the normal values even after two days. Such a slow recovery is to be expected for deep Forbush decreases (Lockwood, 1971; Lagoida, Voronov, and Mikhailov, 2019), but in this case, it was prolonged even more because of the influence of a series of interplanetary disturbances.

This FE could have been registered with an even greater magnitude, but in the morning of May 11, a GLE was observed (related to an X5.8 solar flare at 01:10 UT). This GLE was assigned number 74 in the IGLED database (https://gle.oulu.fi, Usoskin et al., 2020), it had an amplitude of ~8% (according to SOPO NM). Based on the GSM results its amplitude may be estimated at around 2%. This GLE is slightly visible in the GSM data on CR density in Figure 6 in the form of a small bump right after the FE minimum. It was discussed multiple times at various conferences, and some works were already published on it. For example Mavromichalaki et al. (2024) had presented results of the NKUA "GLE Alert++ system": some NMs (LMKS, OULU, SOPO) entered the alert status in several minutes and a notification was issued on May 11, 2024 at 02:05 UT. Also Chilingarian et al. (2024) discussed various parameters of this GLE. In particular, energy releases were estimated based on the integral count rates of two ground detectors. Hayakawa et al. (2024) briefly described GLE 74, providing ground-based NM records from high-latitude regions (OULU, THUL, SOPO, MWSN). It is important to note that GLE is a rare phenomenon, yet we have already observed GLE occurrences in large (>10%) FEs. As such, only on June 11, 1991 (an FE with 21.6% magnitude) the GLE 51 was recorded during its decrease phase, and in other cases GLEs occurred during the phase of recovery (2-3 days after the FE minimum): GLE 7, GLE 24, GLE 44, GLE 51.

As mentioned, an extreme geomagnetic storm was observed during May 10-11. Such a strong disturbance of the magnetic field clearly had to influence the measurements of ground detectors on Earth. So-called magnetospheric effect was observed during FEs many times before and is described in literature (Baisultanova, Belov, and Yanke, 1995; Belov et al., 2005; Ghag et al., 2023). Figure 6 compares CR density variation obtained with the GSM with (brown curve) and without (blue curve) magnetospheric effect correction. The correction is performed based on hourly values of the Dst-index (see Belov et al., 2015). The difference between the resulting curves reaches 4% as seen in Figure 6.

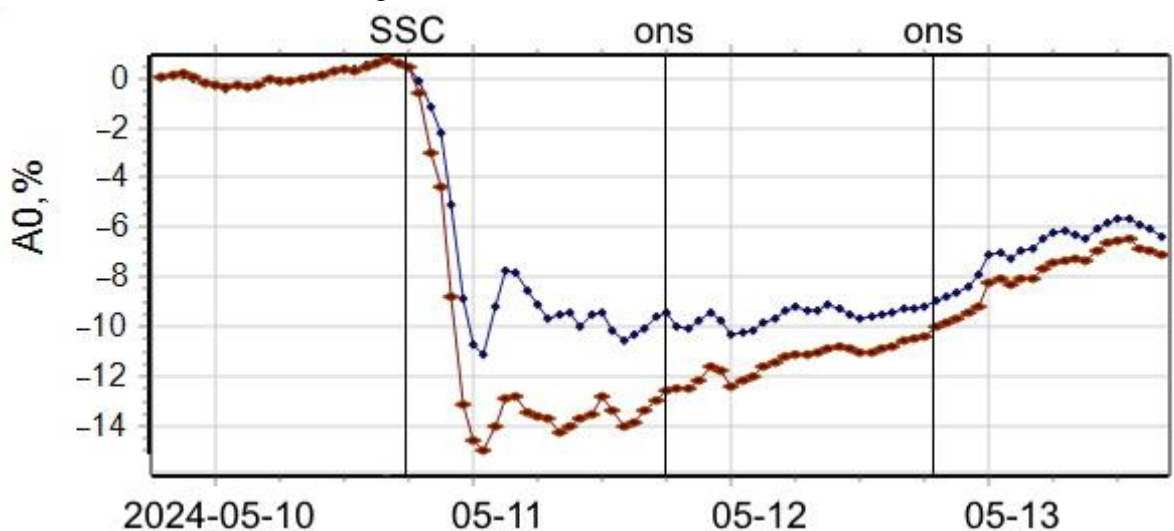

Figure 6. CR density computed with the GSM for 10 GV particles with (brown curve) and without (blue curve) magnetospheric effect correction.

The presence of a magnetic cloud (MC) structure in this event is up for debate. For example it is not present in the catalog of SW structures maintained at Space Research Institute (http://www.iki.rssi.ru/omni/catalog/2024/20240422c.jpg), but Hajra et al. (2024) describe three periods with MC structures according to per-minute SW data: from ~03:28 to 07:54 UT, from ~08:25 to 10:36 UT, and from ~11:19 to 17:17 UT on May 11. Yet CR variations, and in particular anisotropy variations, may often give additional hints about the inner structure of interplanetary disturbances. An abrupt change in the direction of the equatorial component of CR anisotropy was observed at around 09:00 UT, May 11 (see coupled vector diagram, Fig. 5). There were also a change of the sign of the north-south component of CR anisotropy (thin blue bars attached to the A0 curve, the scale is the same as for A0) at ~11:00 UT and then reverse change at ~17:00 UT, together with the MC characteristic parameters in SW data: low plasma-$\beta$, elevated IMF values without discontinuities or waves (see, e.g., Burlaga et al., 1981; Klein and Burlaga, 1982). All the described features lead us to believe some distinguishable structure was present in this interplanetary disturbance, that was large enough to influence high-energy CR. We suppose that despite the complex nature of interplanetary disturbance, the MC structure was somewhat preserved and passed the Earth from 11:00 UT to 17:00 UT, May 11 (marked by yellow stripes in Figs 4-5).

## 4. Conclusion

The paper considers and discusses in detail the changes in the parameters of solar activity, interplanetary medium, geomagnetic activity as well as variations in the cosmic ray density and anisotropy (calculated by the global survey method for particles with a rigidity of 10 GV), in the period from May 8 to May 13, 2024. During this time, powerful solar flares and fast CMEs were recorded, leading to registration of an extreme geomagnetic storm along with one of the most significant Forbush effects for the entire observation period.

The FE under study was one of the most outstanding in several parameters: the magnitude of CR density variations (15.7%), the maximum hourly (4.4%) and two-hour decrement (8.8%) of CR density. But in terms of the values of the equatorial and north-south CR anisotropy components, on the contrary, it had the smallest values compared to other large (> 10%) FEs.

One of the interesting features of the FE in question is the registration of GLE 74 during this event. Although only a small enhancement was observed, the combination of "large FE + GLE inside" is quite rare.

During the interplanetary disturbances on May 10-13, an extreme geomagnetic storm was recorded, due to which a significant magnetospheric effect was observed in the NM data (~4%).

According to the authors' opinion, based on SW and CR data (in particular, on the behavior of CR anisotropy) one may conclude that an MC was present in the studied event on May 11 lasting from 11:00 UT to 17:00 UT.

**Acknowledgments:** The authors would like to thank the Neutron Monitor Database (NMDB) for kindly providing cosmic ray data. The authors are grateful to all the solar, geomagnetic and interplanetary data providers.

**Conflicts of Interest:** The authors declare that there is no conflict of interest.